\documentclass[fleqn,10pt]{wlscirep}
\usepackage[T1]{fontenc}

\title{Numerical Modeling of Oxygen Diffusion in Tissue Spheroids Undergoing Fusion Using Function Representation and Finite Volumes}

\author[1]{Katherine Vilinski-Mazur}
\author[1,2*]{Bogdan Kirillov}
\author[1,3]{Oleg Rogozin}
\author[1]{Dmitry Kolomenskiy}

\affil[1]{Center for Materials Technologies, Skolkovo Institute of Science and Technology, Moscow, Russia}
\affil[2]{Center for Precision Genome Editing and Genetic Technologies for Biomedicine, Institute of Gene Biology, Russian Academy of Sciences, Moscow, Russia}
\affil[3]{Federal Research Center "Computer Science and Control", Russian Academy of Sciences, Moscow, Russia}

\affil[*]{Bogdan.Kirillov@skoltech.ru}

\usepackage{amsmath,amssymb}
\usepackage{changepage}
\usepackage{textcomp,marvosym}
\usepackage{cite}
\usepackage{nameref}
\usepackage[right]{lineno}
\usepackage{microtype}
\DisableLigatures[f]{encoding = *, family = * }
\usepackage{array}
\newcolumntype{+}{!{\vrule width 2pt}}
\newlength\savedwidth

\raggedright
\usepackage[aboveskip=1pt,labelfont=bf,labelsep=period,justification=raggedright,singlelinecheck=off]{caption}

\makeatletter
\renewcommand{\@biblabel}[1]{\quad#1.}
\makeatother

\usepackage{lastpage,fancyhdr,graphicx}
\usepackage{epstopdf}
\fancyheadoffset[L]{2.25in}
\fancyfootoffset[L]{2.25in}
\usepackage{soul}
\usepackage{url}
\usepackage{graphicx}
\usepackage{amsmath}
\usepackage{booktabs}
\usepackage{xcolor}
\usepackage{siunitx}
\usepackage{easyeqn}

\begin{abstract}
  A three-dimensional cell culture called a spheroid serves as a foundational entity in a wide variety of modern tissue engineering applications, including 3D-bioprinting and preclinical drug testing. Lack of oxygen within tissue spheroids hinders metabolism of cells and eventually leads to cell death. Prevention of necrosis is crucial to success of tissue engineering methods and such prevention requires estimation of cell viability in the spheroid. We propose a novel approach for numerical modeling of diffusion in tissue spheroids during their fusion. The approach is based on numerical solutions of partial differential equations and the application of Function Representations (FRep) framework for geometric modeling. We present modeling of oxygen diffusion based on meshes derived from the geometry of fusing spheroids, a method for selecting optimal spheroid size, and several statistics for estimating cellular viability. Our findings provide insights into oxygen diffusion in three-dimensional cell cultures thus improving the robustness of biotechnological methods that employ tissue spheroids.
\end{abstract}

\newcommand{\V}{\Omega}
\newcommand{\dV}{\partial \V}
\newcommand{\intCell}{\int_{\V}}
\newcommand{\intFaces}{\oint_{\dV}}
\newcommand{\implicit}[1]{\underbrace{#1}_\text{implicit}}
\newcommand{\explicit}[1]{\underbrace{#1}_\text{explicit}}

\date{}

\begin{document}

\flushbottom
\maketitle
\thispagestyle{empty}

\section*{Introduction}
The demand for donor organs was steadily increasing for the last twenty years due to several factors that include an increase in success rate of transplantation surgeries and the growing rates of vital organ failures~\cite{caplan2016finding} linked with changes in lifestyle, aging population and prevalence of chronic diseases~\cite{levitt2015could}. That increase in demand lead to a dramatic shortage in supply of transplantable organs, with hundreds of thousands patients on a waiting list and an average of seventeen people dying every day due to the inaccessibility of suitable transplants (in USA, according to Health Resources and Service Administration~\cite{organdonorOrganDonation}). 

Most currently implemented remedies~\cite{abouna2008organ} for organ shortage include merely a change of policies: switch from opt-in to opt-out system for donation, monetary support for a family of deceased or living donor, educational programs to raise awareness and so on. While undeniably important, these options do not provide an actionable solution for an underlying problem (lack of organs) and act rather as a form of support for the healthcare system.  

From technological point of view, there exists several possible angles of attack for organ shortage, such as various organ preservation techniques~\cite{maathuis2007perspectives, hameed2017advances}, fast transportation systems~\cite{sage2022testing}, organ-donor matching algorithms~\cite{duquesnoy2008clinical, mattei2017mechanisms} and artificial organs~\cite{kolff2002artificial}. Note that while these technologies are well-conceptualized, most are still in the research stage and have years and years ahead before any kind of adoption.
The current paper is dedicated to a refinement for one of such technologies, 3D bioptinting. We propose an improved modeling of oxygen diffusion processes in the printed construct. 

Currently under active development, 3D bioprinting holds promise for solving organ shortage by providing means to fabricate organs on demand~\cite{jovic20203d, panda2022focused} while the other aforementioned technologies solve problems of transplantology infrastructure (fast transportation, organ-donor matching, preservation). Bioengineered organs~\cite{atala2000tissue}, created by combining cells and scaffolds, have the potential to replace damaged or non-functional organs entirely, while by utilizing spheroids as the starting material, 3D bioprinting can create more complex and functional tissue structures compared to using a one-dimensional cellular culture comprised of individual cells within a hydrogel matrix~\cite{robu2019using}. Compared to production of artificial organs, 3D bioprinting does not have a problem of integrating constructs made of synthetic materials into living tissue -- there is no steep change in mechanical properties that introduce mechanical stress~\cite{zhang2023optimization} and no immune response (in case of using the patient's primary cell culture~\cite{derman2023high}).

Typical pipeline of bioprinting with spheroids consists of four stages. During the first stage, the spheroids are produced in a laboratory setting using methods like ``hanging drop"\cite{mehesz2011scalable}, forced aggregation~\cite{baraniak2012scaffold} or microfluidics~\cite{vadivelu2017microfluidic}. Then comes the stage of bioink preparation~\cite{sun2022simple}. Bioink consists of cells in form of spheroids and a hydrogel that provides structural function. Hydrogel allows for the spheroids to be precisely put at correct coordinates during the third stage, deposition of biological material, which involves the layered addition of spheroids and non-linked hydrogel followed by the initiation of gel crosslinking~\cite{daly2021bioprinting}. After bioprinting, the bioconstruct undergoes the fourth stage, maturation~\cite{datta2018essential, shafiee2017cellular}. Under osmotic pressure, the spheroids start to fuse together and the cells placed in an incubator start to divide. Spheroids form functional tissue by merging with each other. 

Main distinction between natural and bioprinted tissue lies within the organization of supply of oxygen and nutrients – natural tissue is vascularized while biofabrication of blood vessels is itself a task not yet solved (we suggest the works of Mir et al.~\cite{mir20233d} and Chae et al.~\cite{chae20233d} for a thorough review for vascularization problem in 3D biofabrication). Therefore, most studies of 3D bioprinting use avascular spheroids and bioprinted constructs. There are a number of approaches to solve vascularization (for example, microtubes~\cite{chen2023microtube} or induced angiogenesis~\cite{mazloomnejad2023angiogenesis}), but the current study focuses on avascular spheroids made from normal, non-tumorous, cells.

During the lifecycle of a non-vascularized spheroid, there is little to no new oxygen present right before the last stage of the aforementioned pipeline. Lack of oxygen leads to the emergence of hypoxic region, followed by necrosis and then degradation of a bioprinted construct~\cite{cui20173d, wang20213d, apelgren2021long}. Modeling the oxygen distribution can provide insights into dynamics of the system and help estimate the time that is left for printing and maturation. With modeling, we can select the optimal size of a spheroid which would maximize the time of normal operation for a bioprinted construct. 

Oxygen distribution for normal tissue spheroids in bioprinting is particularly challenging because most of the research is done (and thus most of experimental data is known) for tumors~\cite{leek2016methods} (there are notable exceptions though, for example the works of Sego et al.~\cite{sego2017heuristic}). Tumor spheroids have many useful applications themselves, serving as an~\textit{in vitro} model for chemotherapy selection~\cite{ma2012multicellular, song2022single, hu2022vascularized, perche2012cancer} and drug design~\cite{roy2023tumor, wu2008microfluidic, lazzari2017multicellular}, but this research is focused on the fusion and necrotic region formation of \textit{normal} tissue in context of application for bioprinting. Question of whether the tumor studies are transferable for normal tissues is not that clear either, albeit these two biological systems are comparable. Malignant tumor cells and cells from a normal tissue are not the same with regards to their physiology in aspects such as cell cycle regulation~\cite{williams2012cell}, proliferation~\cite{bresciani1968cell} and metabolic activity~\cite{deberardinis2016fundamentals}. These fundamental differences necessitate a reevaluation of the methodologies and conclusions drawn from studies focused on pathological tissues.

Distribution of oxygen in multicellular spheroids is described by a reaction-diffusion equation that models the diffusion of oxygen over the volume of the spheroid, inflow and outflow of oxygen through the boundaries and consumption of oxygen by the cells. There are analytical solutions for oxygen partial pressure for spherical, oblate and prolate shape~\cite{grimes2018oxygen} of spheroids but these solutions consider the boundary rather smooth, while in reality it consists of deformed cells that experience different amounts of pressure from their surroundings. Effect of surface irregularities cannot be easily inferred using the works of Grimes et al.~\cite{grimes2018oxygen}, because they do not assume the existence of these irregularities.

To model such variable geometry, we use Function Representation (FReps)~\cite{pasko1995function, pasko2011procedural} is a method for modeling geometry with noise. To describe the random deformations in solid geometry we use Gardner noise~\cite{gardner1984simulation}. This road leaves construction of analytical solution for the reaction-diffusion equations impractical, so we opt for a numerical solution, which we get with Finite Volume Method (FVM)~\cite{eymard2000finite, mazumder2015numerical}. The application of FReps allows us to model not only a single spheroid, but the fusion of two spheroids geometrically, so we can construct a mesh for a numerical method on any stage of spheroid fusion. 

Currently, the distribution of oxygen is usually included as a part of hierarchical schemes using lattice-based discrete models, such as Cellular Potts (for example, the works of Jiang et al.~\cite{jiang2005multiscale}, Li et al.~\cite{li2014effects} and Poplawski et al.~\cite{poplawski2009front} regarding avascular tumors, works of Sego et al.~\cite{sego2017heuristic} regarding non-tumorous spheroids and works of, again, Sego et al.~\cite{sego2020unification} and Bustamante et al.~\cite{bustamante2021biofabrication} regarding biofabrication), agent-based modeling, such as PhysiCell~\cite{ghaffarizadeh2018physicell}, and other cellular automata-like approaches. 

However, these models simulate the growth of a spheroidal configuration, which is a much slower process than the diffusion of oxygen, and also much more complex, and those models are computationally expensive. Here, we analyze the distribution of oxygen over synthetic biologically plausible shapes without assuming or modeling their growth. This simple approach allows focusing on the geometrical effect of the surface boundary.

Our main contributions can be summarized as follows:
\begin{enumerate}
	\item We suggest a combination of geometrical modeling with Function Representation and Finite Volume Method to solve reaction-diffusion equations. Our aim is to take aberrations from spherical shape and smooth surface into account;
	\item For the first time, we use the combination of FReps and FVM to model the distribution of oxygen during the fusion of two spheroids; 
	\item We perform a comprehensive computational study for influence of surface deformities on the oxygen diffusion;
    \item We model dependency of hypoxia on spheroid diameter and determine the range of sizes at which hypoxic regions form, allowing for selection of acceptable diameter for optimal oxygen supply.
\end{enumerate}

\section*{Results}

\subsection*{Combination of FRep and FVM provides a robust numerical method for analysis of diffusion in multicellular tissue spheroids}

\begin{figure*}[h]
    \centering
    \includegraphics[width=\textwidth]{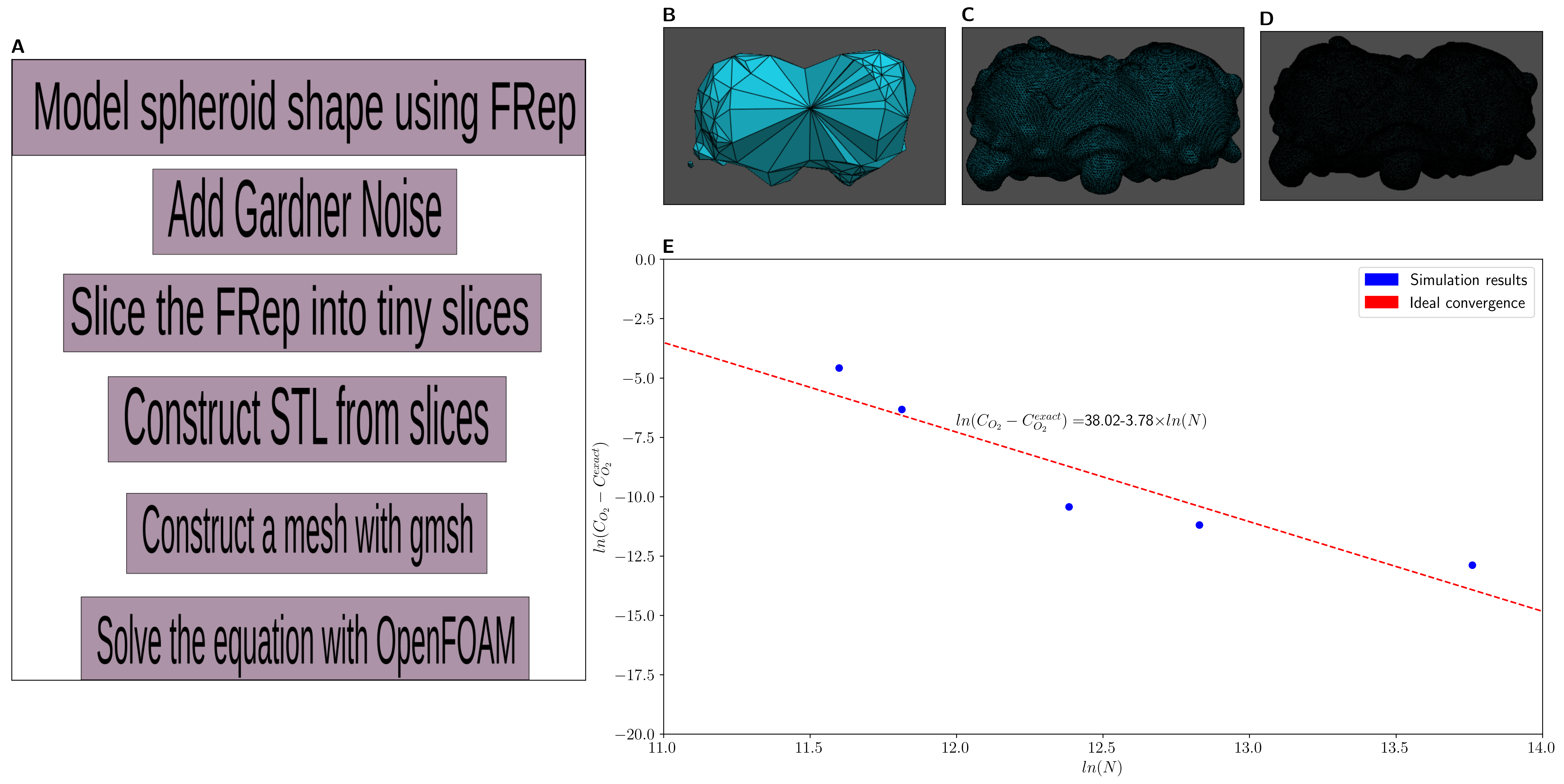}
    \caption{Pipeline visualization and convergence plots for FRep+FVM method. (A) Description of our pipeline. (B) Coarse mesh (344 polygons). (C) Medium mesh (\num{109000} polygons). (D) Fine mesh (\num{980000} polygons). (E) Convergence plot for oxygen partial pressure---the equation shows the trend line with the coefficients found by linear regression.}
    \label{fig:figure1}
\end{figure*}

We combined FRep (Function Representation) and FVM (Finite Volume Method) to analyze the diffusion in multicellular tissue spheroids. Roughly, our pipeline consists of six steps: modeling of a spheroid with FRep as a sphere, adding solid noise to the surface of the spheroid, slicing the FRep models into tiny slices, constructing the STL (stereolithgraphy format) file from the slices, building the mesh out of the STL file, running FVM simulation on the mesh. A schematic overview of the pipeline is shown in Figure\ref{fig:figure1}A.

The pipeline we develop starts with Function Representation and introduces variability through Gardner noise, then creates an STL file and a mesh for detailed numerical analysis. The pipeline steps, outlined at Figure~\ref{fig:figure1}A, encapsulate a comprehensive methodology for modeling, simulating, and analyzing diffusion processes in complex geometry of the tissue spheroid. 

Modeling of spheroid shape using Function Representation, the first step, is performed using the computer-assisted design application FRepCAM~\cite{popov2021cad}. Gardner noise addition, the second step, as described in the Materials and Methods section, was already implemented as the primitive operation in FRepCAM CAD software. 

The third step of the pipeline, slicing~\cite{nayyeri2022planar} of three-dimensional geometric models, FRep models in particular, is the process of dividing a three-dimensional object into a series of two-dimensional cross-sections or slices. This technique is commonly used in computer-aided design (CAD) and computer graphics to create physical prototypes or visual representations of complex objects. 

Slicing allows for the visualization of internal structures, measurements, and other details that may not be apparent in a three-dimensional view. It is also used in additive manufacturing processes such as 3D printing, where each slice is printed layer by layer to build the final object. To slice the FRep model, the work of Maltsev et al.~\cite{maltsev2021accelerated} was used. 

Slicing consists of the following steps:
\begin{enumerate}
    \item Constructing the 2D bounding boxes for all the unit cells of the FRep model;
    \item Building the spatial index structure with the precalculated bounding boxes of the unit cells in each layer of the sliced 3D model;
    \item Applying the compound adaptive criterion~\cite{maltsev2021accelerated} based on the spatial query during quadtree construction for the FRep model;
    \item Applying the acceleration criterion~\cite{maltsev2021accelerated} based on the spatial search during calculating the defining function at every point of interest in the space;
    \item Creating the topology of the curve using the marching squares (MS~\cite{hanisch2004marching}) algorithm and the connected component labelling (CCL~\cite{samet1981connected}) algorithm;
    \item Calculating the exact values of the implicit curve on the edges of adjacent cells using numerical methods for solving nonlinear equations.
\end{enumerate}
The output of slicing is then used to construct an STL file, the fourth step of the pipeline~(Figure ~\ref{fig:figure1}A), by taking the boundary coordinates of the slices and forming the required list of vertices and triangles. After that, Gmsh~\cite{geuzaine2009gmsh} constructs the mesh for OpenFOAM from the STL file, and OpenFOAM software is used for solving the diffusion equation. We have used default parameters everywhere except for the mesh size, which we varied.

 The result of this approach is a description for the distribution of oxygen over the volume of the spheroid. We have identified the specific parameters that need to be tuned at each step of the pipeline to achieve an optimal mesh and ensure a robust solution for the reaction-diffusion equation. Using the FRep framework allows us to model the fusion of spheroids using elementary operations within the set of FRep modifiers: blending and union (the results are shown in Figures~\ref{fig:figure1}B,C,D and~\ref{fig:figure2}A,B,C).

When determining optimal mesh density for robust estimation of oxygen distribution and capturing precise geometry details, several factors come into play, like distribution of mesh elements, shape of mesh element, method of mesh generation and number of elements. Since we are using unstructured mesh, the distribution of elements depends on the method of mesh generation as well as the shapes of individual elements. For our experiments, we use the Gmsh~\cite{geuzaine2009gmsh} software for generation of tetrahedral volume meshes based on the surface meshes derived from the STL files. Studying the influence of meshing algorithms on the results is outside the scope of the current research. Increasing number of elements generally improves resemblance to actual behavior being modelled. However, there is a trade-off between accuracy and efficiency, as analyzing an infinite number of elements is not feasible---very coarse meshes (small number of elements) tend to obscure fine geometry details. Striking the right balance is crucial. 

We have analyzed behavior of FVM solutions on coarse (Figure~\ref{fig:figure1}B), medium (Figure~\ref{fig:figure1}C) and fine meshes (Figure~\ref{fig:figure1}D) while varying number of polygons in the mesh. According to Figure~\ref{fig:figure1}E, we have reached acceptable convergence in all cases (the points scatter near optimal convergence shown as red dashed line). We attribute the deviations in convergence plot to the fact that our mesh is unstructured. In the context of computational fluid dynamics and finite volume methods, structured mesh configurations, constructed of regularly arranged uniform elements, usually exhibit superior convergence characteristics and resolution capabilities compared to unstructured meshes due to their rigid connectivity. However, construction of structured mesh limits our ability to accurately represent complex FRep-derived geometry we use due to their lack of flexibility in geometry representation~\cite{zhu2013finite}, therefore we choose to sacrifice easy convergence for flexibility and completeness of geometry representation.

\subsubsection*{Gardner Noise allows for inclusion of surface irregularities into numerical simulation}

We have performed several computational experiments to show the effect of spheroid surface irregularities on oxygen distribution over its volume. We have used Gardner Noise to model surface deformities of a spheroid. Gardner Noise is controlled by three parameters (Equation~\eqref{eq:gardner}): amplitude ($a$), phase ($p$), and frequency ($q$). The parameters are discussed further in the ``Materials and Methods'' section.

We need to separate studying the effect of surface irregularities on the convergence of the diffusion equation, on cellular viability and on the dynamic of oxygen distribution:
\begin{enumerate}
    \item Convergence. The first entity, the effect on convergence, studied in previous section, is considered insignificant (as shown in Fig.~\ref{fig:figure1}E). 

    \item Viability. The second entity, effect on viability, cannot be directly inferred from our simulation but is known to be significant (see the review of Han et al.~\cite{han2021challenges} for reference). We hypothesize that we can infer the effect of irregularities on cell survival using distribution of oxygen as a proxy.

\end{enumerate}

To study the effect of surface irregularities on distribution of oxygen, first, we fixed the number of polygons in the mesh for further experiments involving varying solid noise level. An extreme amount of noise may require a larger mesh to cover adequately without losing fine details. To provide sufficient space for any considered noise levels, we set our polygon number at \num{300000}.

Then, we applied linear regression on log-transformed minimal partial pressures from the first 10 time points to analyze the evolution of oxygen distribution. The parameters obtained from this linear regression ($\hat{y}=A+Bx$) represent our statistics that characterise the results of the simulation. By focusing on the initial 10 time points, we aim to model the rapid changes in oxygen distribution rather than studying steady-state results of later time frames. In all cases, the time interval of the numerical simulation was sufficiently long to reach a steady distribution of oxygen partial pressure. Also note that the spheroid size was held constant during each oxygen diffusion simulation. We performed Ordinary Least Squares test via statmodels library~\cite{seabold2010statsmodels} and Automatic Relevance Determination (ARD~\cite{mackay1994bayesian}) via scikit-learn~\cite{pedregosa2011scikit} to study the effiects of amplitude, phase and frequency on $A$ and $B$ and corrected for two comparisons using Bonferroni correction, thus setting our p-value to 0.025 for OLS. ARD does not require computation of p-values for interpretation, we should instead interpret the size of regression coefficients. ARD forces the final model to be sparse and pushes the coefficients for insignificant variables to be zero. 

\begin{figure*}[h]
    \centering
    \includegraphics[width=\textwidth]{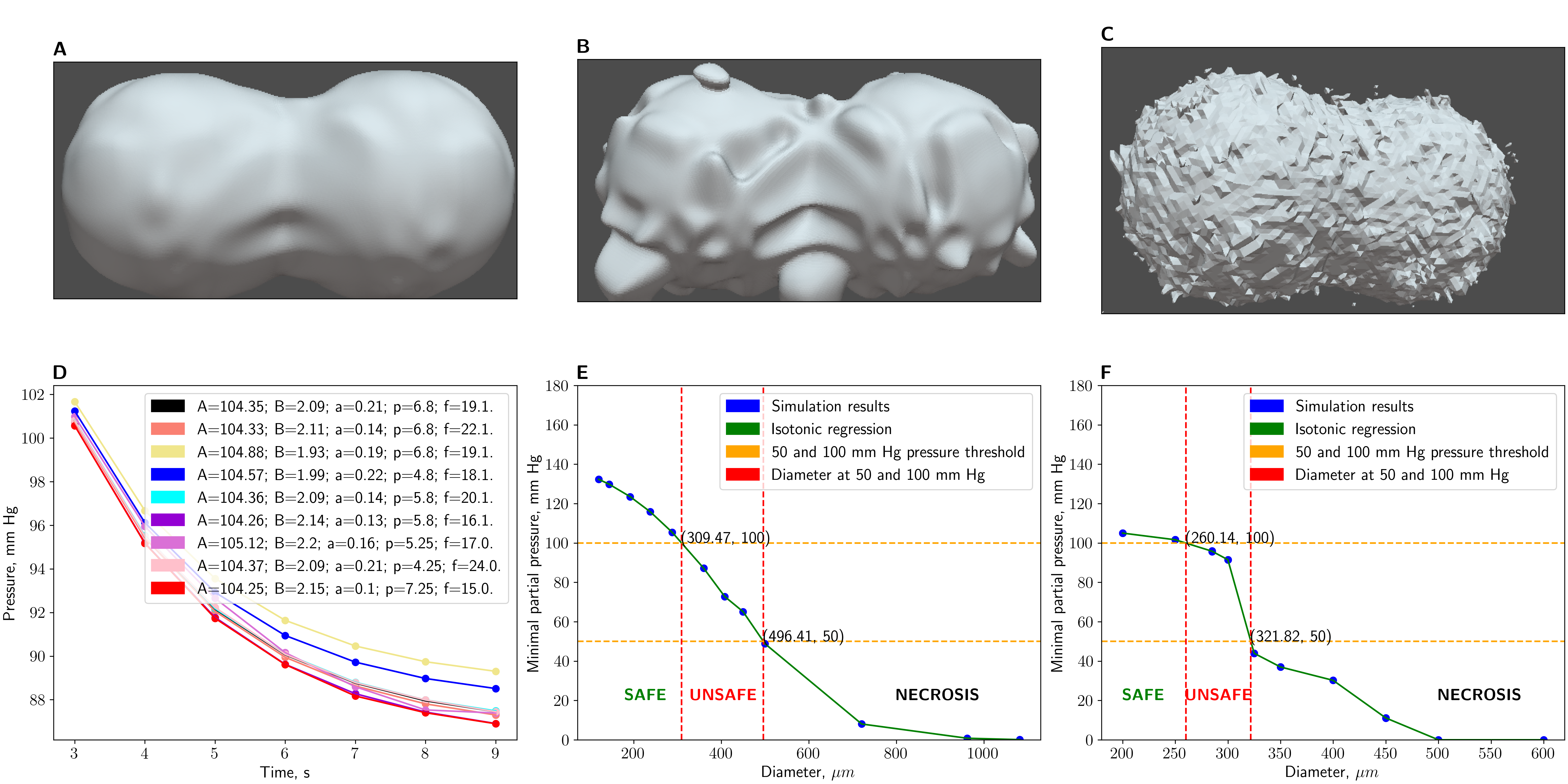}
    \caption{Study of Gardner Noise influence on distribution of oxygen and optimization of spheroid diameter for minimization of hypoxic region size. (A) Fusion of two spheroids with noise of small amplitude and phase. (B) Fusion of two spheroids with noise of large amplitude. (C) Fusion of two spheroids with noise of high frequency. (D) Evolution of partial pressure during time conditional on amplitude, phase and frequency. (E) Dependency of minimal partial pressure on diameter of spheroid for a single spheroid. Safe and unsafe diameter ranges are indicated by the labels. (F) Dependency of minimal partial pressure on diameter of spheroid for two fused spheroids. Safe and unsafe diameter ranges are indicated by the labels.}
    \label{fig:figure2}
\end{figure*}

According to OLS, the only significant ($p = 0.013$ for $A$, non-significant for $B$) dependency of oxygen distribution dynamic described with coefficients of linear regression on parameters of Gardner Noise is dependency on phase, but this dependency is only borderline significant because it does not pass stricter p-value thresholds and according to Figure~\ref{fig:figure2}D, there is only slight dissociation of the time-P($O_2$) plots. Moreover, many values of phase produce very unrealistic images of the surface (e.g. impossibly sharp peaks, separations of small parts from the main spheroid volume), which Figure~\ref{fig:figure2}C shows. Automatic Relevance Determination, on the other hand, showed slight negative dependency of $B$ on amplitude, with the following linear models: $\hat{A} = 0 \times a + 0 \times q + 0 \times p + 104.5$ and $\hat{B} = -0.82 \times a + 0 \times q + 0 \times p + 2.22$. Both these models output nearly constant values for $A$ and $B$ respectively, therefore, we are bound to conclude that ARD did not find any learnable connection between amplitude, phase, frequency and coefficients that describe oxygen dynamics over time. This implies that surface deformities do not measurably affect the oxygen diffusion, which is an interesting development because the spheroid shape irregularities are known to affect cellular viability~\cite{kosheleva2016laser}. This discrepancy with prior experimental works~\cite{kosheleva2016laser} and simulations~\cite{poplawski2009front} could be explained by the fact that our approach is intended to model already grown spheroids and does not consider neither circumstances of their development nor defects during spheroid production. However, it is crucial to note that while our study focuses on mature spheroids, the initial stages of spheroid formation and early development phases are indeed critical periods where surface deformities, discrepancies in environmental conditions and introduced defects could significantly impact cellular viability and function. Future research (including the continuation of previously discussed Cellular Potts works) should aim to simulate the mechanisms underlying this adaptability and study them experimentally, as well as explore potential interventions that could mitigate the effects of defects and experimental errors during these sensitive developmental stages. Moreover, the discrepancy observed between our findings and prior experimental and computational works underscores the importance of considering the life cycle of spheroids in research design.

\subsubsection*{Mathematical modeling uncovers the dependency of hypoxia on spheroid size}
We aimed to model dynamics of hypoxic region formation considering the size of spheroid. This would allow us to determine the ``safe'' range of diameters that are unlikely to have hypoxic regions inside as well as ``unsafe'' ranges where hypoxic regions start to form or are already formed and dominate the distribution of oxygen (shown on Figure~\ref{fig:figure2}E and F). 
The diffusion equation was solved in the interior of the spheroids. The initial condition was the constant and uniform partial pressure distribution of oxygen. The boundary condition was the constant and uniform oxygen partial pressure on the boundary. We have analyzed the dependence of \(pO_{2}\) on diameter for the case of one spheroid and two fused spheroids with the following parameters, initial and boundary conditions:
\begin{enumerate}
	\item Diffusion coefficient in cells~\cite{grimes2014oxygen}: \(D = \SI{2e-9}{\m\squared\per\s}\);
	\item Initial and boundary partial pressure of oxygen in DMEM~\cite{Foncesa2018}: $P_{O_2}^{DMEM} = \SI{140}{\mmHg}$;
	\item Constant oxygen consumption rate~\cite{grimes2014oxygen}: $OCR=\SI{20}{\mmHg\per\s}$.
\end{enumerate}

According to Figure~\ref{fig:figure2}E and F, the trend of oxygen minimal partial pressure dropping as the diameter of the spheroid grows is evident. 

We define hypoxic region as a region that has oxygen partial pressure lower than a survival threshold. According to the work of Mueller--Klieser and Sutherland~\cite{mueller1982oxygen}, the first signs of necrosis for EMT6/Ro and V-79-171B tumor cultures emerge at \(pO_{2}\) of \SI{57}{\mmHg} and \SI{42}{\mmHg} respectively. Gomes et al.~\cite{mueller1982oxygen} have shown experimentally that 5\% (\SI{38}{\mmHg} at room temperature and 1 atm pressure) oxygen concentration extremely hinders the growth of spheroids of HCT116 human colon adenocarcinoma cell line. 

We consider that the survival threshold for \textit{normal}, non-tumorous cell lines would be significantly higher than aforementioned values because tumor cells can adapt to oxygen deprivation (for example, by expressing Hypoxia-Inducible Factors~\cite{choudhry2018advances} that lead to angiogenesis~\cite{chen2023hypoxic}). We therefore consider a threshold range between 50 mmHg and 100 mmHg.

The first threshold indicates the beginning of change in the dynamics that we seek. We hypothesize that in real spheroids, the dynamics we model in computational experiments describes the formation of hypoxic regions, hence the last threshold indicates its completion. These points determine the range of ongoing hypoxia formation: an optimally large spheroid (the one without hypoxia) would have the minimal oxygen partial pressure greater than 100 mmHg, in the "safe" range, as Figure~\ref{fig:figure2} (panels E and F) shows. Spheroids that have less than 50 mmHg minimal oxygen partial pressure are in necrosis. According to the aforementioned figure, 50 mmHg is achieved at \SI{309.5}{\um} and \SI{260}{\um} for single and two fused spheroids, respectively; therefore, we conclude that optimal spheroids would be smaller than that – 250-\SI{300}{\um} and 220-\SI{250}{\um} respectively. The selection of the exact spheroid size is left to the experimenter.

\section*{Discussion}
This work is based on application of geometric modeling and finite volume methods for studying the diffusion of oxygen in multi-cellular tissue spheroids. The chosen approach for geometric modeling, Function Representation (FRep), allowed us to carefully include the explicitly parameterized description of surface irregularities into the simulation. Thus, our model provides a straightforward and easy way to study their effect on distribution of oxygen.

The meshes obtained from FRep descriptions of spheroid geometry were put to work in a Finite Volume Method setup, allowing us to solve diffusion equations numerically. We were able to investigate a wide variety of possible surface deformities, from almost smooth sphere to a sphere with very sharp peaks and use this information to understand how the lifelike irregularities of the surface affect the distribution of oxygen and, thus, viability of the cells. 

Contrary to our initial expectations, our analysis revealed that the irregularities on the spheroid surface had negligible effects on the overall distribution of oxygen. These findings challenge the notion that surface deformities significantly impact oxygen distribution and cellular viability in multi-cellular tissue spheroids~\cite{zanoni20163d}. Further research is needed to fully comprehend the underlying mechanisms and to explore additional factors that contribute to the observed robustness of oxygen distribution in the presence of surface deformities.

Nevertheless, in this study, we were able to find out \textit{in silica} the largest size of a spheroid without necrosis for bovine chondrocytes. When the spheroid exceeds this size, the necrotic region emerges. Therefore, tissue spheroids made of bovine chondrocytes of size \SI{288}{\um} and above are not applicable for 3D bioprinting experiments. Our recommendation for size selection is to pick smaller spheroids that do not reach the hypoxia formation size range, so a value slightly below \SI{288}{\um} would suffice, for example something in the range of 250-\SI{280}{\um} and in case of fused spheroids to pick a size below \SI{200}{\um}, for example, 170-\SI{190}{\um}.

Our work is similar to the works of Grimes et al.~\cite{grimes2014oxygen, grimes2014method} in its aim and setting, but we have made significant changes in the modeling approach. We were working with normal spheroids (bovine chondrocytes) while the aforementioned papers considered tumor-derived ones and we have studied complex spheroid geometries instead of less-lifelike spherical and elliptic models. For the first time, we have modeled the distribution of oxygen during the fusion of two spheroids within the context of fusion geometry. The modeling of oxygen distribution during the fusion of two spheroids has indeed been previously explored~\cite{poplawski2009front}, however, our approach uniquely incorporates explicit geometric considerations into the modeling process, offering a novel perspective on how the spatial arrangement and morphological changes of fusing spheroids influence oxygen dynamics. 

We have shown in simulation a known trend of \(pO_{2}\) decrease with increasing size~\cite{mueller1982oxygen}, and we observe a plateau starting at \SI{500}{\um} with oxygen partial pressure nearby 0 for the case of two fused spheroids. Considering the single spheroid case, the plateau starts further to the right, near \SI{900}{\um}, however at \SI{500}{\um}, the partial pressure reaches previously established hypoxia start diameters for tumor cells (according to Mueller et al\cite{mueller1982oxygen}, the first signs of necrosis for EMT6/Ro and V-79-171B tumor cultures emerge at \(pO_{2}\) of \SI{57}{\mmHg} and \SI{42}{\mmHg} respectively). Moreover, \SI{500}{\um}, according to the work of Vinci et al~\cite{vinci2012advances}, is the threshold value for emergence of necrosis in tumor spheroids. Thus, the results of our simulation for a single spheroid converge with the existing literature. 

 Our future research endeavors primarily focus on the experimental validation of the fusion model proposed in the current study. The experimental validation would provide empirical evidence to support the effectiveness and applicability of our modeling approach.

We opt to use a non-vascularized tissue spheroid for our experiments due to our specific focus on describing the diffusion process within the cellular culture. The modeling of vascularized spheroids holds significant importance~\cite{kwak2020vitro} due to their ability to replicate properties of solid tumors, including cell-cell and cell-ECM interactions~\cite{han2021challenges}. These characteristics make vascularized spheroids valuable for studying drug resistance, radiation resistance, cancer cell migration, invasion, and angiogenesis~\cite{brassard2020vitro}. While studying vascularized spheroids is fundamental for tissue engineering~\cite{figtree2017vascularized}, it introduces additional complexity, that is not necessary for our specific research objectives. It requires considering computational fluid dynamics of blood flow within the vessel, while for non-vascularized spheroid, it is sufficient to consider only the diffusion.

Following the example of Grimes et al.~\cite{grimes2014method, grimes2014oxygen, grimes2018oxygen}, we use the distribution of oxygen as a proxy for cellular viability because it reflects how much oxygen can be used in the metabolism of the cell. This approach is standard in the field, however, there are two significant research gaps between studying cellular viability and oxygen consumption. The first gap follows from direct experimental measurement of oxygen distribution in the cell being hard to perform, as most methods provide only a point measurement of oxygen consumption rate (e.g. Clark electrodes~\cite{brito2021oxygen, clark1987personalized}, Respiratory Detection System~\cite{strovas2010direct}) which is not enough to reconstruct the distribution over the volume of the spheroid. Other methods require techniques like Positron Emission Tomography~\cite{mirabello2018oxygen, muehllehner2006positron}, Optical Coherence Tomography~\cite{huang2017optical} or Paramagnetic Resonance Oximetry~\cite{langan2016direct} which are not easily accessible in most laboratories. A second gap in the research is the immense difficulty of viability prediction from the first principles, as the bulk of the research (including the current work) focuses on mathematical modeling of different viability proxies, like oxygen or nutrient consumption~\cite{han2021challenges}, but we have yet to see any method to predict actual cell survival measured by Live/Dead assay or a similar approach. We have found out, that contrary to the assumptions of other research groups, the surface irregularities have a negligible effect on the distribution of oxygen, but their effect on cellular viability is well known and replicated in a number of works~\cite{zanoni20163d, de2018high, amaral2017comparative, kosheleva2016laser}. We observe a certain decoupling between cellular viability which is affected by the deformities negatively and its proxy, oxygen distribution, which remains very similar to the elliptic and spherical cases. The reason why oxygen distribution is not a perfect map of cell survival is still to be discovered. It is essential to recognize that the relationship between oxygen distribution and cell survival is influenced by a multitude of factors beyond just oxygen availability. These include the distribution of nutrients, the presence of signaling molecules, and the overall environmental conditions surrounding the spheroids. While our study sheds light into important aspects of oxygen dynamics, it also underscores the complexity of cellular responses within three-dimensional structures like spheroids. Therefore, future investigations should aim to integrate a holistic view of the microenvironment, including oxygen and nutrient gradients, to fully understand the determinants of cell survival and functionality within spheroids. 
While Cellular Potts models offer valuable insights into multicellular dynamics and tissue organization, their limitations (oversimplification of cellular dynamics, discreteness, dependency on hyperparameters, lack of molecular details) highlight the importance of integrating complementary modeling approaches and experimental validation to achieve a comprehensive understanding of biological systems. Bridging these research gaps is a promising direction for future studies requiring the involvement of multidisciplinary teams consisting of biologists, physicists, chemists, and machine learning specialists. Nevertheless, studying oxygen distribution in fully grown multicellular tissue spheroids supplies significant information about cellular viability and contributes to our understanding of cell behavior in a realistic environment.

\section*{Materials and Methods}

\subsection*{Geometrical modeling with Function Representations}

The Function Representation (FRep) framework defines a geometrical object using a real-valued continuous function:
\begin{EQA}[c]\label{eq:best}
	f(x, y, z),
\end{EQA}
where $f: R^3 \rightarrow R$. The function $f$ has positive values inside the object, negative
values outside, and zero on the surface~\cite{pasko2010procedural} and the object boundary $f(x, y, z) = 0$ is named ``implicit surface". The FRep approach is a method that provides a heterogeneous representation of objects of high geometrical complexity (e.g. a mammalian cell colony \cite{savchenko1995simulation, savchenko1999computer}). Such functions may be used to define any arbitrary shape of particles and model topological changes. 
In the current work, we apply Function Representation for modeling of three-dimensional cellular cultures. To do so, we first define a spherical shape with FRep, then we apply Gardner noise to model surface irregularities, and after that, we define \emph{blending} and \emph{union} operations to model smooth fusion of two spheroids. The completed model is then converted into an STL file and processed by a mesh construction program (we use Gmsh~\cite{geuzaine2009gmsh}).

\subsection*{Modeling surface irregularities with Gardner noise}
To model the natural surface of the tissue spheroid, the Gardner noise (GN) function is used~\cite{pakhomova2021modeling}:
\begin{EQA}[c]
  GN(x,y,z,a,q,p) =
  \left(a \sin(qx) + \frac{a}{1.17}\sin\left({\frac{qx}{1.35} + p \sin(qz)}\right)\right)
  \times \\
  \left(a \sin(q y) + \frac{a}{1.17} \sin\left(\frac{qy}{1.35} + p \sin(qx)\right)\right) \times \\
  \left(a\sin(qz) + \frac{a}{1.17}\sin\left(\ \frac{qz}{1.35} + p \sin(qy)\right)\right),
  \label{eq:gardner}
\end{EQA}
where \(x,y,z\) are coordinates of a point on a surface, \(a,p,q\) are amplitude, phase, and frequency of Gardner noise, respectively. The frequency determines the number of surface deformities occurring on a model. The amplitude of Gardner noise controls the average height of each deformity. The phase determines the shape of each deformity, whether it is more smooth or more cube-like. Higher frequencies result in more frequent deformities, larger amplitudes result in more pronounced deformities, and different values of phase alter the shape of the deformity. 

\subsection*{Blending and union operations for modeling the spheroid fusion}
To model the fusion of spheroids, we use \emph{blending} operations coupled with \emph{union} operation~\cite{pasko2011procedural}. We use a mathematical framework of Rvachev functions (R-functions~\cite{rvachev1963analytical}) to define \emph{blending} and \emph{union}. Rvachev functions allow for transition between FReps that define geometrical figures and Boolean functions. By introducing R-functions, we can construct logical operations over shapes.
R-functions that we use are as follows:
\begin{EQA}[c]
		F\left( f_{1},f_{2} \right) = R\left( f_{1},f_{2} \right) + d\left( f_{1},f_{2} \right) 
	\label{eq:rvachev}
\end{EQA}
where \(R (f_1, f_2)\) is the R-function corresponding to the \emph{union blending} operation, \(f_1\) and \(f_2\) are defining functions of two initial tissue spheroids, \(d\left( f_{1},f_{2} \right)\) is the Gaussian-type displacement function, \(a_o\), \(a_1\) and \(a_2\) are parameters controlling the shape of the blend.
The displacement function is defined as follows~\cite{pasko2002bounded}:
\begin{EQA}[c]
	d(f_1,f_2) = \frac{a_0}{1+(f_1/a_1)^2+(f_2/a_2)^2}
\end{EQA}
Following the works of the Pasko group~\cite{pasko2004space}, we use the following R-function for union:
\begin{EQA}[c]
	R_{un}(f_1,f_2) = f_1 + f_2 + \sqrt{f_1^2 + f_2^2},
\end{EQA}
and the corresponding R-function~\cite{pasko2010procedural} for blending union:
\begin{EQA}[c]
	F_{b.un} = f_1 + f_2 + \sqrt{f_1^2 + f_2^2} + \frac{a_0}{1+(f_1/a_1)^2+(f_2/a_2)^2}.
\end{EQA}

\subsection*{Numerical solution for diffusion and consumption of oxygen}

Oxygen partial pressure is an integral component regulating developmental processes, cell metabolism, and functioning of tissue~\cite{reis2019encyclopedia}. \textit{In vivo}, tissues experience a wide range of oxygen partial pressures, depending on their location and capillary supply, which are notably different from the inhaled oxygen partial pressure of about 20\%. To describe the evolution of oxygen partial pressure, we use a diffusion-reaction equation with a source term describing oxygen consumption. The oxygen partial pressure $P$ is proportional to the oxygen concentration, which changes due to diffusion and consumption by the cells. Fick's law describes a diffusive flux:
\begin{EQA}[c]   
	I = - D \nabla P,
	\label{fick}
\end{EQA}
where $D$ is the coefficient of diffusion, which is determined by the mobility of molecules.
Combining together with the constant rate oxygen consumption $OCR$, we obtain the following differential equation:
\begin{EQA}[c]
	\frac{\partial P(\vec{r},t)}{\partial t} = D\Delta P(\vec{r},t) - OCR,
	\label{consume}
\end{EQA}
where $\Delta$ is the Laplace operator. At the boundaries of the computational domain, we set the Dirichlet boundary condition.

The numerical solution of~\eqref{consume} is obtained in  OpenFOAM~\cite{openfoamOpenFOAMFree}, a widely used open-source toolkit for Computational Fluid Dynamics (CFD). OpenFOAM provides an efficient implementation of complex physical models in continuum mechanics, mainly based on the Finite Volume Method (FVM), which involves dividing the computational domain into finite control volumes and then approximating the differential equations in integral form over these volumes. The differential equation~\eqref{consume} is rewritten in the following integral form suitable for FVM:
\begin{EQA}[c]
    \frac{\partial}{\partial t}\intCell P \,dV =
    \implicit{\intFaces D\frac{\partial P}{\partial\vec{r}} \cdot\vec{n}\,dS} -
    \explicit{ \intCell OCR \,dV},
 \label{eq:num_mth}
\end{EQA}
where $\V$ is the control volume, $\vec{n}$ is the unit vector normal to its boundary $\dV$. Equation~\eqref{eq:num_mth} is discretized such that the diffusion term is accounted for implicitly. In our simulations, we use
the Euler method to integrate~\eqref{eq:num_mth} over time. 
Numerical fluxes are calculated by the Gauss linear limited Laplacian scheme. This scheme ensures stability and accuracy for rather irregular meshes.

\subsection*{Linear models for analysis of surface irregularity influence on oxygen partial pressure}
To describe the influence of surface irregularities (modeled by Gardner noise), we focus on the linear part of the time-pressure dependency and approximate it with a simple linear model:
\begin{EQA}[c]
	\hat{P}_{O_2}(t) = A + B t,
	\label{eq:linear}
\end{EQA}
where $A$ and $B$ are the model parameters that act as test statistics for oxygen pressure field evolution. We assume that if the influence of irregularities is significant, then the coefficients of the linear model would vary significantly. We analyze this variability with Automatic Relevance Determination and Ordinary Least Squares.
Automatic Relevance Determination is a regression method that produces sparse coefficients. Each coefficient is drawn from zero-mean Gaussian distribution:
\begin{EQA}[c]
	p(w|\alpha) = \prod_{i=0}^N N(w_i|0, \alpha_i^{-1}),
\end{EQA}
with $\alpha$ vector of N + 1 hyperparameters~\cite{tipping2001sparse}. ARD assigns different \textit{relevance weights} $\alpha$ to features and models these \textit{relevance weights} as latent features.
Ordinary Least Squares allow for the selection of parameters for linear models that minimize the sums of squared differences between the regression labels and the output of the linear model from Equation~\eqref{eq:linear}. We use test for coefficients being equal to zero to determine whether the modeled dependency is statistically significant and our \textit{p-value} threshold is determined using Bonferroni correction:
\begin{EQA}[c]
	p_{new} = \frac{p_{old}}{N_{comparisons}}.
\end{EQA}

\subsection*{Isotonic regression}
To identify thresholds based on the time--pressure curve, we need a mathematical method that could be used to interpolate the dependency between observed points. We know that there is both a hard lower limit on the partial pressure of oxygen (\SI{0}{\mmHg}), and a hard upper limit, which is equal to the partial pressure of oxygen in the atmosphere (\SI{160}{\mmHg} at sea level); therefore, we would need to consider these limits in our predictive model. To do that, we use isotonic regression \cite{barlow1972isotonic, luss2014generalized}, a non-parametric regression method that requires no assumptions about the shape of the approximated function other than its monotonicity. It solves the following optimization problem:
\begin{EQA}
	min \sum_{i} w_i (y_i - \hat{y}_i)^2,
\end{EQA}
where \(X, y, w \in R\) are input data, labels and weights respectively. The solution should obey the following condition: 
\begin{EQA}
	\forall i,j \in I: \hat{y}_i \leq \hat{y}_j \iff \ X_i \leq X_j.
\end{EQA}

\section*{Author contributions}
K.V.M., B.K., O.R. and D.K. developed all concepts and designed the study. K.V.M. performed all computational experiments. B.K. performed statistical analysis of the results. K.V.M., B.K. and D.K. wrote the main manuscript text. D.K. supervised the research. O.R. supervised the diffusion equation study. All authors have read, reviewed and approved the manuscript.

\section*{Competing Interests}
The authors declare no competing interests.

\section*{Funding Information}
This work was supported by the Skoltech Translational Research and Innovation Program grant "Spheroid Revolution". B.K. is supported by the Ministry of Science and Higher Education of the Russian Federation [075-15-2019-1661].

\section*{Data availability}
Data sets generated during the current study are available from the corresponding author on reasonable request.

\bibliography{references}
\end{document}